\documentclass[aps,nofootinbib,twocolumn,prd,eqsecnum,showpacs,showkeys,preprintnumbers]{revtex4-1}

\usepackage[caption=false]{subfig}
\usepackage{graphicx}
\usepackage{amsmath}
\usepackage{amsfonts}
\usepackage{amssymb}
\usepackage{color}
\usepackage{bm}
\usepackage{mathrsfs}
\usepackage{epstopdf}
\usepackage{url}
\usepackage{footnote}
\usepackage{textcomp}

\makeatletter
\newcommand*{\rom}[1]{\expandafter\@slowromancap\romannumeral #1@}
\makeatother

\begin{document}

\title{Tradeoff between Smoother and Sooner ``Little Rip''}
\author{Mariam Bouhmadi-L\'{o}pez $^{1,2}$}
\email{mariam.bouhmadi@ehu.es}
\author{Pisin Chen $^{3,4,5,6}$}
\email{chen@slac.stanford.edu}
\author{Yen-Wei Liu $^{3,5}$}
\email{f97222009@ntu.edu.tw}
\date{\today}

\affiliation{
${}^1$Department of Theoretical Physics, University of the Basque Country
UPV/EHU, P.O. Box 644, 48080 Bilbao, Spain\\
${}^2$IKERBASQUE, Basque Foundation for Science, 48011, Bilbao, Spain\\
${}^3$Department of Physics, National Taiwan University, Taipei, Taiwan 10617\\
${}^4$Graduate Institute of Astrophysics, National Taiwan University, Taipei, Taiwan 10617\\
${}^5$Leung Center for Cosmology and Particle Astrophysics, National Taiwan University, Taipei, Taiwan 10617\\
${}^6$Kavli Institute for Particle Astrophysics and Cosmology, SLAC National Accelerator Laboratory, Stanford University, Stanford, CA 94305, U.S.A.
}

\begin{abstract}
There exists dark energy models that predict the occurrence of ``little rip''. At the point of little rip the Hubble rate and its cosmic time derivative approach infinity, which is quite similar to the big rip singularity except that the former happens at infinite future while the latter at a finite cosmic time; both events happen in the future and at high energies. In the case of the big rip,  a combination of ultra-violet and infra-red effects can smooth its doomsday. We therefore wonder if the little rip can also be smoothed in a similar way. We address the ultra-violet and infra-red effects in general relativity through a brane-world model with a Gauss-Bonnet term in the bulk and an induced gravity term on the brane. We find that the little rip is transformed in this case into a sudden singularity, or a ``big brake". Even though the big brake is smoother than the little rip in that the Hubble rate is finite at the event, the trade-off is that it takes place sooner, at a finite cosmic time. In our estimate, the big brake would happen at roughly 1300Gyr.
\end{abstract}

%%%%%%%%%%%%%%%%%%%%%%%%%%%%%%%%%%%%%%%%%%%%%%%%%%%%%%%%%%%%%%%%%%%%%%%%%%%%%%%%%

\maketitle

\section{introduction}\label{introduction}

Recent observational evidences show that the expansion of our universe is currently accelerating \cite{SNa,WMAP,SDSS}. This discovery has opened several possibilities to explain the nature of the late-time acceleration. By constructing reasonable models that account for this phenomenon, the nature of the cosmic acceleration could be either attributed to an exotic dark energy component that makes up roughly 70 percent of the universe \cite{Copeland:2006wr} , or described by a modified gravity theory such that the predictions are consistent with the current observational data \cite{Clifton:2011jh}. If the late-time accelerated expansion is driven by dark energy, the simplest setup is by adding a cosmological constant $\Lambda$ to Einstein equation with cold dark matter on top of the baryonic sector ($\Lambda$CDM model), which fits very well the observations even though the cosmological constant has to be extremely fine-tuned as compared with the expected value from quantum field theory \cite{Weinberg:1988cp,Carroll:2000fy}.

However, there are several other candidates that can play the role of dark energy, which can be characterized by an equation of state $w$. The equation of state $w$ of a given matter content is defined by the ratio between its pressure and its energy density, i.e., the equation of state $w\equiv P/\rho$, where $P$ and $\rho$ are the pressure and the energy density of the matter, respectively. From current observations, the equation of state of dark energy is very close to $-1$ \cite{SNa,WMAP,SDSS}. Nevertheless, unlike the cosmological constant whose equation of state is strictly equal to $-1$, the equation of state of dark energy could deviate from $-1$ a bit and vary with time. Furthermore, dark energy with equation of state $w<-1$ has not been excluded observationally \cite{SNa,WMAP,SDSS}. which is also known as phantom energy \citep{Caldwell:1999ew}. If this hypothetical phantom energy exists in the late-time universe, our universe will enter into a super-accelerating epoch and space-time singularities might occur in the future. Within a Friedmann-Lema\^{\i}tre-Robertson-Walker (FLRW) metric, dark energy related future space-time singularities can be classified as follows depending on different types of divergence \cite{Nojiri:2005sx,Nojiri:2008fk,Bamba:2008ut,footnote1}:
\begin{itemize}
\item Big rip singularity: the singularity happens at a finite cosmic time $t$ where the scale factor $a$, the Hubble parameter $H$ and its cosmic time derivative $\dot{H}$ diverge \cite{Caldwell:1999ew,Starobinsky:1999yw,Carroll:2003st,Caldwell:2003vq,Chimento:2003qy,Dabrowski:2003jm,
    GonzalezDiaz:2003rf,GonzalezDiaz:2004vq,Nojiri:2004pf}, i.e., when $t\rightarrow t_s$, $a\rightarrow\infty$, $H\rightarrow\infty$ and $\dot{H}\rightarrow\infty$;

\item Sudden singularity: the singularity happens at a finite cosmic time with a finite scale factor, where the Hubble parameter remains finite but its cosmic time derivative diverges \cite{Barrow:2004xh,Gorini:2003wa,Nojiri:2005sx}, i.e., when $t\rightarrow t_s$, $a\rightarrow a_s$, $H\rightarrow H_s$ and $\dot{H}\rightarrow\infty$;

\item Big freeze singularity: the singularity happens at a finite cosmic time with a finite scale factor where the Hubble parameter and its cosmic time derivative diverge \cite{BouhmadiLopez:2006fu,BouhmadiLopez:2007qb,Nojiri:2005sx,Nojiri:2004pf,Nojiri:2005sr}, i.e., when $t\rightarrow t_s$, $a\rightarrow a_s$, $H\rightarrow \infty$ and $\dot{H}\rightarrow\infty$;

\item Type \rom{4} singularity: the singularity happens at a finite cosmic time with finite scale factor where the Hubble parameter and its cosmic time derivative remain finite, but higher cosmic time derivatives of the Hubble parameter still diverge \cite{Nojiri:2005sx,Nojiri:2008fk,Bamba:2008ut};
\end{itemize}
where $t_s$, $a_s$ and $H_s$ are finite constants. However, in addition to the possible space-time singularities mentioned above and connected with dark energy, there exist other possible abrupt events that are smoother than those enumerated above and are also connected with dark energy:
\begin{itemize}
\item Little rip event: there is no future space-time singularity; however, the scale factor , the Hubble parameter and its cosmic time derivative all diverge at an infinite cosmic time \cite{Nojiri:2005sx,Nojiri:2005sr,Ruzmaikina1970,Stefancic:2004kb,BouhmadiLopez:2005gk,Frampton:2011sp,Brevik:2011mm,Frampton:2011rh,Nojiri:2011kd}, i.e., when $t\rightarrow\infty$, $a\rightarrow\infty$, $H\rightarrow\infty$ and $\dot{H}\rightarrow\infty$. This kind of behavior was first found in Ref.~\cite{Ruzmaikina1970}.

\item $w$-singularity: the equation of state $w$ diverges at a finite cosmic time with a finite scale factor, where the energy density $\rho$ and the pressure $P$ vanish \cite{Dabrowski:2009kg,FernandezJambrina:2010ps}, i.e., when $t\rightarrow t_s$, $a\rightarrow a_s$, $\rho\rightarrow0$, $P\rightarrow0$ and $w\rightarrow\infty$;

\item Pseudo-rip event: the Hubble parameter approaches a constant as the cosmic time goes to infinity with an infinite scale factor, where the disintegration of bound structures may or may not occur in this case \cite{Frampton:2011aa}, i.e., when $t\rightarrow\infty$, $a\rightarrow\infty$ and $H\rightarrow H_s$.
\end{itemize}

Among these events, the little rip is quite similar to a big rip singularity except that the former happens at infinite future while the latter at a finite cosmic time. Such event, despite avoiding a future singularity at a finite cosmic time, will still lead to the destruction of all structures in the universe. Additionally, the little rip has previously been analyzed under four-dimensional (4D) standard cosmology \cite{Nojiri:2005sx,Nojiri:2005sr,Ruzmaikina1970,Stefancic:2004kb} in a dilatonic brane-world model \cite{BouhmadiLopez:2005gk}. Recently, this event has been named ``little rip'' \cite{Frampton:2011sp,Brevik:2011mm,Frampton:2011rh,Nojiri:2011kd,Belkacemi:2011zk}. In addition, the difference between the big rip and little rip events becomes conventional when UV quantum gravitational effects are taken into account, since the Planck curvature is reached in a finite cosmic time in both kinds of events.

If our universe is filled with phantom dark energy, a doomsday can happen at high energies in the far future. These possible future behaviors have motivated the interest in searching for resolutions where the abrupt events can at least be appeased, and several works on possible future space-time singularity avoidance have been carried out (see for example the Refs.~\cite{BouhmadiLopez:2004me,Elizalde:2004mq,Abdalla:2004sw,Dabrowski:2006dd,Kamenshchik:2007zj,BouhmadiLopez:2009pu,
Sami:2006wj,BouhmadiLopez:2005gk,BouhmadiLopez:2008bk,Setare:2008mb}). In the case of a big rip, a combination of ultra-violet (UV) and infra-red (IR) modifications to general relativity (GR) can smooth its doomsday \cite{BouhmadiLopez:2009jk}. We therefore wonder if the little rip event can also be smoothed in a similar way, since it takes place in a high-energy regime where UV effect is required, and in a late-time universe where IR effect is required. Here we follow the similar procedure to address UV and IR effects in GR through a spatially flat 4D brane embedded in a 5D anti-de-Sitter (AdS$_5$) bulk that is modified by both Gauss-Bonnet (GB) and induced gravity (IG) terms with the matter fields confined on the brane, where the GB correction manifests in the high-energy regime while the IG effect becomes significant in the late-time universe. For a recent paper on brane-world models with a GB term in the bulk see please Ref.~\cite{Maeda:2011px} and the references therein for an extensive list of references on the GB brane-world scenario.

Furthermore, we focus here on the self-accelerating branch of this model, which contains the Dvali-Gabadadze-Porrati (DGP) self-accelerating solution modified by the GB effect (DGP-GB) \cite{Dvali:2000hr,Dvali:2000xg}, due to the fact that this branch gives both UV and IR modifications to GR. Finally, the investigation on the little rip event can also shed some lights on the nature of such GB and IG modified brane-world system.

The outline of this paper is as follows. In Sec.~\rom{2}, we consider a spatially flat brane-world within an AdS$_5$ bulk with GB and IG modifications, and we review the self-accelerating branch obtained from this model. In Sec.~\rom{3}, we examine if for the same matter content that would induce a little rip event in standard GR can be smoothed in some manner within the framework reviewed in Sec.~\rom{2}. In Sec.~\rom{4}, we analyze the late-time behavior of the DGP-GB model filled with phantom dark energy, and furthermore we estimate when the sudden singularity will take place in this case. Finally, in Sec.~\rom{5}, we summarize our results and present our conclusions.

\section{self-accelerating branch}

The generalized Friedmann equation of a spatially flat brane embedded in an AdS$_5$ bulk with both GB and IG curvature effects is given by \cite{Kofinas:2003rz,footnote3}:
\begin{align}
&\left[1+\frac83\alpha\left(H^2-\frac{\mu^2}2\right)\right]\sqrt{H^2+\mu^2}\notag\\
&=\pm \,r_c\left[\frac{\kappa^2_4}{3}\left(\rho+\lambda\right)-\gamma H^2\right]\label{Friedmann},
\end{align}
where the parameter $\mu^2$ is defined by
\begin{equation}
\mu^2=\frac1{4\alpha}\left(1-\sqrt{1+\frac43\alpha\Lambda_5}\right),
\end{equation}
$\alpha(\geq0)$ is the GB parameter which has the dimension of length square, $\gamma$ is a dimensionless parameter controlling the strength of the IG term, $\Lambda_5(\leq0)$ is the bulk cosmological constant, $\lambda$ is the brane tension and the crossover scale $r_c$ is given as $r_c\equiv\kappa_5^2/2\kappa_4^2$. Therefore the parameter $\mu^2$ is bounded as $0\leq\mu^2<1/4\alpha$ (see also Ref.~\cite{footnote2}).

The branch with ``$+$'' sign in Eq.~(\ref{Friedmann}) contains the DGP normal branch solution (with $\Lambda_5=0$, $\alpha=0$, $\lambda=0$ and $\gamma=1$); while the ``$-$'' sign in Eq.(\ref{Friedmann}) contains the DGP self-accelerating solution. Here we focus on the self-accelerating branch with ``$-$'' sign that gives the UV and IR modifications to GR.
%%%%%%%%%%%%%%%%%%%%%%%%%%%%%%%%%
\begin{table*}[!ht]
\begin{tabular}{|c|c|l|}
\hline
$b$  & $\bar\rho_1$ and $\bar\rho_2$ &  \hspace{8em} Solutions for $\bar X$         \\
\hline
&  & $\bar X_1=-1/3\left[2\sqrt{1-3b}\cosh{(\eta/3)}-1\right]$; $0<\bar{\rho}_1<\bar{\rho}$        \\
&  & $\bar X_1=-1/3\left[2\sqrt{1-3b}\cos{(\theta/3)}-1\right]$; $0<\bar{\rho}<\bar{\rho}_1$    \\
&  & $\bar X_2=1/3\left[2\sqrt{1-3b}\cos{((\pi+\theta)/3)}+1\right]$; $0<\bar{\rho}<\bar{\rho}_1$    \\
&  & $\bar X_3=1/3\left[2\sqrt{1-3b}\cos{((\pi-\theta)/3)}+1\right]$; $0<\bar{\rho}<\bar{\rho}_1$        \\
$0\leq b<1/4$ &$\bar\rho_2\leq0<\bar\rho_1$  & \textbf{Limiting case:}\\
&  & $\bar{X}_1=-1/3\left[2\sqrt{1-3b}-1\right]$; $0<\bar\rho_1=\bar\rho$\\
&  & $\bar{X}_2=\bar{X}_3=1/3\left[\sqrt{1-3b}+1\right]$; $0<\bar\rho_1=\bar\rho$\\
&  & $\bar{X}_1=\bar{X}_2=\left.-1/3\left[\sqrt{1-3b}-1\right]\right|_{b=0}=0$; $0=\bar\rho_2=\bar\rho$\\
&  & $\bar{X}_3=\left.1/3\left[2\sqrt{1-3b}+1\right]\right|_{b=0}=1$; $0=\bar\rho_2=\bar\rho$\\
\hline
&  & $\bar X_1=-1/3\left[2\sqrt{1-3b}\cosh{(\eta/3)}\right]$; $0<\bar\rho$               \\
$1/4\leq b<1/3$  &$\bar\rho_2<\bar\rho_1\leq0$  & \textbf{Limiting case:} \\
&  & $\bar{X}_1=\left.-1/3\left[2\sqrt{1-3b}-1\right]\right|_{b=1/4}=0$; $0=\bar{\rho}_1=\bar{\rho}$\\
&  & $\bar{X}_2=\bar{X}_3=\left.1/3\left[\sqrt{1-3b}+1\right]\right|_{b=1/4}=1/2$; $0=\bar{\rho}_1=\bar{\rho}$\\
\hline
$b=1/3$  & $\bar\rho_1=\bar\rho_2=-1/27$ & $\bar X_1=-1/3\left[(1+27\bar\rho)^{1/3}-1\right]$  \\
\hline
$1/3<b$  & $\bar\rho_1$, $\bar\rho_2$: complex conjugates & $\bar X_1=-1/3\left[2\sqrt{3b-1}\sinh{(\xi/3)}\right]$
\\
\hline
\end{tabular}
\caption{Exact solutions of the modified Friedmann equation (\ref{cubic}). These solutions are divided into four different regions depending on the value of the parameter $b$, and $\bar X_1$, $\bar X_2$, and $\bar X_3$ correspond to three possible branches in different ranges of the parameter $b$. Here we restrict to the physical solutions where the total dimensionless energy density $\bar \rho$ is positive.}
\label{solutions}
\end{table*}
%%%%%%%%%%%%%%%%%%%%%%%%%%%%%%%%%
To solve the Eq.(\ref{Friedmann}) analytically, it is more convenient to introduce the following dimensionless parameters \cite{BouhmadiLopez:2008nf,BouhmadiLopez:2009jk,BouhmadiLopez:2011xi,Belkacemi:2011zk,BouhmadiLopez:2012uf}:
\begin{align}
\bar{X}=&\frac{8}3\frac{\alpha}{\gamma r_c}\sqrt{H^2+\mu^2}\label{barX},\\
b=&\frac{8}3\frac{\alpha}{\gamma^2 r_c^2}\left(1-4\alpha\mu^2\right)\label{b},\\
\bar{\rho}=&\frac{64}9\frac{\alpha^2}{\gamma^2 r_c^2}\mu^2+\frac{32}{27}\frac{\alpha^2\kappa^2_5}{\gamma^3r_c^3}\left(\rho+\lambda\right)\label{rhobar}.
\end{align}
Then the Friedmann equation (\ref{Friedmann}) can be rewritten on a simpler form:
\begin{equation}
\bar{X}^3-\bar{X}^2+b\bar{X}+\bar{\rho}=0\label{cubic}.
\end{equation}
Notice that we restrict to ``$-$'' sign of Eq.(\ref{Friedmann}), and $\bar X$ is positive [cf.~Eq.(\ref{barX})]. Therefore only the positive solutions of Eq.(\ref{cubic}) are physically meaningful.

The number of real solutions of the cubic equation (\ref{cubic}) depends on the sign of the discriminant $N$ \cite{Abramowitz}:
\begin{equation}
N=Q^3+R^2,
\end{equation}
where
\begin{equation}
\left\{
\begin{array}{ll}
\displaystyle Q=\frac13\left(b-\frac13\right),\\
\displaystyle R=\frac16b+\frac12\bar{\rho}-\frac1{27}\,\,.
\end{array}
\right.
\end{equation}
It is more convenient to factorize the discriminant $N$ as
\begin{equation}
N=Q^3+R^2=\frac14(\bar{\rho}-\bar{\rho}_1)(\bar{\rho}-\bar{\rho}_2),
\label{discriminant}
\end{equation}
where
\begin{align}
\bar{\rho}_1&=-\frac13\left\{b-\frac29\left[1+\sqrt{(1-3b)^3}\right]\right\}, \label{rho1}\\ \bar{\rho}_2&=-\frac13\left\{b-\frac29\left[1-\sqrt{(1-3b)^3}\right]\right\}.
\label{rho2}
\end{align}
If $N$ is positive, there is only one real solution. If $N$ is negative, all the solutions are real, and if $N$ vanishes, all the solutions are real and at least two of them are equal.

\begin{figure}[h]
\centering
  \includegraphics[width=6.4cm]{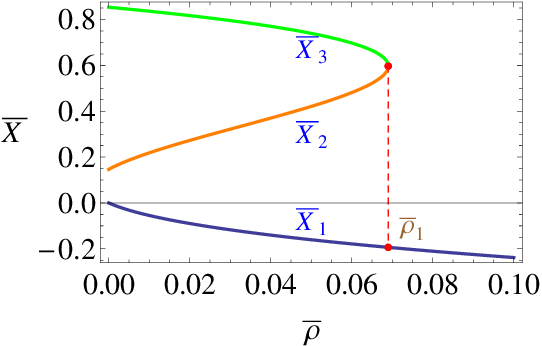}
	\caption{Branches of the modified Friedmann equation (\ref{cubic}). We plot $\bar X$, essentially the Hubble rate [see Eq.(\ref{barX})], versus the dimensionless energy density $\bar\rho$. For this plot, we set $b=1/8$ as an example. Notice that only positive $\bar X$ is allowed [see Eq.(\ref{barX})], and the solution $\bar X_2$ corresponds to the self-accelerating branch.}
	\label{branches}
\end{figure}

After doing a careful analysis of the cubic Friedmann equation (\ref{cubic}), the analytical solutions are summarized in Table \ref{solutions} (see also Fig.~\ref{branches}), where the parameters $\eta$, $\theta$ and $\xi$ are defined as follows:
\begin{align}
&\cosh{\eta}=\frac{R}{\sqrt{-Q^3}}\,\,,  &&\sinh{\eta}=\sqrt{\frac{Q^3+R^2}{-Q^3}}\,\,, \\
&\cos{\theta}=\frac{R}{\sqrt{-Q^3}}\,\,, &&\sin{\theta}=\sqrt{\frac{Q^3+R^2}{Q^3}}\,\,,\label{theta}\\
&\sinh{\xi}=\frac{R}{\sqrt{Q^3}}\,\,, &&\cosh{\xi}=\sqrt{\frac{Q^3+R^2}{Q^3}}\,\,,
\end{align}
which give the constraints $0<\eta$ as well as $0<\xi$. In addition, since the dimensionless energy density $\bar\rho$ is always positive [see Eq.(\ref{rhobar})], the parameter $\theta$ is bounded as $0<\theta\leq\theta_{\textrm{max}}$, where
\begin{equation}
\theta_{\textrm{max}}=\arccos{\left(\frac{-2+9b}{2\sqrt{(1-3b)^3}}\right)}.\label{range}
\end{equation}

In general there are three possible evolutionary branches for the modified Friedmann equation (\ref{cubic}) as shown on Table \ref{solutions} and Fig.~\ref{branches}, but one of them is a negative normal branch denoted by $\bar X_1$, which is not physically meaningful since $\bar X$ is always positive [see the definition given in Eq.(\ref{barX})]. Among these positive solutions, the branch $\bar X_2$ corresponds to the self-accelerating branch:
\begin{equation}
\bar X_2=\frac13\left[2\sqrt{1-3b}\cos{\left(\frac{\pi+\theta}3\right)}+1\right],\label{selfaccelerating}
\end{equation}
which contains the DGP self-accelerating branch when the parameters $\Lambda_5$, $\alpha$, and $\lambda$ vanish and $\gamma=1$, which gives both UV and IR modifications to GR. Moreover, this self-accelerating branch $\bar X_2$ could serve to smooth the little rip event in standard GR to a sudden singularity, in the sense that the Hubble parameter remains finite when the universe reaches the singularity as we will show in the next section.

\section{Sudden singularity}
%replaces the little rip in the self-accelerating branch}

The late-time acceleration of the universe can be driven by a dark energy component with an equation of state $w\lesssim-1$, i.e., a phantom energy is possible observationally \cite{SNa,WMAP,SDSS}. If our universe is filled with phantom dark energy, this substance would lead to a super-acceleration phase, causing perhaps space-time singularities in the future \cite{Caldwell:1999ew, Starobinsky:1999yw, Carroll:2003st, Caldwell:2003vq, Chimento:2003qy, Dabrowski:2003jm, GonzalezDiaz:2003rf, GonzalezDiaz:2004vq, Nojiri:2004pf, Barrow:2004xh, Gorini:2003wa, Nojiri:2005sr, BouhmadiLopez:2006fu, BouhmadiLopez:2007qb,footnote4}. Some possible space-time singularities as well as smoother abrupt events within a FLRW universe filled with phantom dark energy are enumerated in Sec.~\ref{introduction}.

One of these abrupt events is the little rip that is quite similar to a big rip singularity as the scale factor, the Hubble parameter, and its cosmic time derivative all approach infinity as in the big rip albeit at infinite future. Even though for a little rip there is no space-time singularity occurrence at a finite cosmic time, all bound structures in the universe will be destroyed on its way to the little rip \cite{Frampton:2011sp}.  Within a relativistic model, i.e., a 4D GR setup, a perfect fluid whose equation of state reads:
\begin{equation}
P=-\rho-A\rho^{\frac12}\label{littlerip},
\end{equation}
where $A$ is a positive constant, can induce a future little rip event \cite{Nojiri:2005sx,Nojiri:2005sr,Ruzmaikina1970,Stefancic:2004kb,Frampton:2011sp}. This kind of event was first found in \cite{Nojiri:2005sx,Nojiri:2005sr,Ruzmaikina1970,Stefancic:2004kb} and also on a dilatonic brane-world model \cite{BouhmadiLopez:2005gk}. For the latter case, the author described the event as a big rip singularity postponed to an infinite future cosmic time. Recently, this kind of event has been named ``little rip'' \cite{Frampton:2011sp}.

Instead of a relativistic GR model, here we consider the self-accelerating branch $\bar X_2$ of the modified Friedmann equation (\ref{cubic}), because it involves UV and IR effects which could be applied to smooth the little rip or at least to appease it. We remind at this regard that the little rip happens at very high energies where we expect some UV corrections to GR, and at the same time in the far future where IR modifications to GR are also to be expected. This is the reason why we consider a dark energy component; more precisely, a perfect fluid described by Eq.(\ref{littlerip}), filling the self-accelerating brane on top of dark matter. We will show that a future sudden singularity appears when the dimensionless energy density $\bar\rho$ approaches the finite constant $\bar\rho_1$. Consequently, a sudden singularity will replace the little rip.

Given that we are interested in the late-time evolution of the brane and the possible avoidance of the little rip, we assume that the brane total energy density is made up of non-relativistic matter as well as phantom energy with equation of state (\ref{littlerip}):
\begin{equation}
\rho=\rho_m+\rho_d=\frac{\rho_{m0}}{a^3}+\rho_{d0}\left(\frac{3A}{2\sqrt{\rho_{d0}}}\ln{a}+1\right)^2,\label{total}
\end{equation}
where $\rho_m$ is the energy density of the non-relativistic matter, $\rho_d$ is the conserved energy density of the phantom dark energy (\ref{littlerip}), $\rho_{m0}$ and $\rho_{d0}$ are the current energy density for matter and phantom dark energy, respectively, and the current scale factor $a_0$ is fixed as $a_0=1$.

In order to analyze the asymptotic future behavior of the self-accelerating branch $\bar X_2$, we need to analyze the future behavior of the Hubble rate given by Eq.~(\ref{selfaccelerating}) and its cosmic time derivative. The cosmic time derivative of the Hubble parameter can be obtained either from the time derivative of the branch $\bar X_2$ (\ref{selfaccelerating}) or from the time derivative of the cubic Friedmann equation (\ref{cubic}). Either ways lead us to the result:
\begin{equation}
\dot{H}=\frac{3\kappa_4^2}{\gamma}\frac{\bar X_2\cdot(\rho+P)}{(3\bar X_2-1)^2-(1-3b)}\,\,,
\end{equation}
where we have already used the energy conservation condition for the total matter content confined on the brane:
\begin{equation}
\dot{\rho}+3H(\rho+P)=0.
\end{equation}
 Therefore, taking into account  the total energy density given by Eq.(\ref{total}), the cosmic time derivative of the Hubble parameter in the self-accelerating branch is given by
\begin{equation}
\dot{H}=\frac{3\kappa_4^2}{\gamma}\frac{\bar X_2\cdot(\rho_m-A\rho_d)}{(3\bar X_2-1)^2-(1-3b)}\,\,.\label{Hdot}
\end{equation}
Notice that the denominator of Eq.(\ref{Hdot}) is always negative in the self-accelerating branch solution, which can be checked using the explicit self-accelerating solution $\bar X_2$ (\ref{selfaccelerating}) where the range of the parameter $\theta$ is given in Eq.(\ref{range}). Therefore, the cosmic time derivative of the Hubble parameter will become positive once the phantom dark energy dominates over the non-relativistic matter, more precisely when $\rho_m<A\rho_d$.

%As soon as the phantom dark energy dominates over the non-relativistic matter, the cosmic time derivative of the Hubble parameter will increase with respect to the scale factor (see Eq.(\ref{Hdot})).
When the dimensionless energy density $\bar\rho$ tends to $\bar\rho_1$, the self-accelerating solution $\bar X_2$ approaches  the constant value (see Table \ref{solutions} and Fig.~\ref{branches}):
\begin{equation}
\bar{X}_2=\frac13\left[\sqrt{1-3b}+1\right],\label{limiting}
\end{equation}
consequently, the Hubble parameter reaches the constant value:
\begin{equation}
H^2=\frac1{64}\left(\frac{\gamma r_c}{\alpha}\right)^2\left(\sqrt{1-3b}+1\right)^2-\mu^2.\label{limitingH}
\end{equation}
From Eqs.(\ref{Hdot})-(\ref{limitingH}) we find that the cosmic time derivative of the Hubble parameter blows up when $\bar\rho\rightarrow\bar\rho_1$, while the Hubble parameter, the energy density and pressure on the brane remain finite at this point. Besides, the cosmic time ellapsed since the present time $t_0$, till the singularity ocurrence at $t_s$ is given by the integral $t_s-t_0=\int_1^{a_s}da/aH$, which is finite since $aH$ is a non-vanishing  analytical function of the scale factor $a$ between the two limits $a_s$ and $a_0=1$, i.e., a sudden singularity takes place at $a_s$. Consequently, a phantom dark energy perfect fluid (\ref{littlerip}) that induces a little rip event in GR is replaced by a sudden singularity, or a ``big brake", through the self-accelerating branch of the brane-world with GB and IG effects. Notice that this event is different from the one found in Refs.~\cite{Shtanov:2002ek,BouhmadiLopez:2010vi}. Besides, the presence of the sudden singularity is a consequence of the combination of an IG term on the brane and a GB term in the bulk (where the bulk is maximally symmetric corresponding to AdS$_5$ or Minkowski), which is different from the branch singularity in a GB brane-world with a negative mass parameter (or a dark radiation term) discussed in Ref.~\cite{Maeda:2011px}. The bulk we have considered is an AdS$_5$ one where there is no branch singularity.

\section{Phenomenology and constraints of the DGP-GB model with ``little rip'' phantom matter}

\subsection{The theoretical framework}
In this section, we consider a spatially flat DGP brane embedded in a Minkowski bulk with a GB modification (DGP-GB), i.e., described by the modified Friedmann equation (\ref{Friedmann}) with a vanishing bulk cosmological constant and brane tension, and $\gamma=1$. The Hubble rate of the branch containing the DGP self-accelerating branch in this case can be rewritten as
\begin{equation}
H^2=\frac{\kappa_4^2}3\rho+\frac1{r_c}\left(1+\frac38\alpha H^2\right)H.\label{DGPGB}
\end{equation}
The current expansion of the universe is dominated by non-relativistic matter and dark energy which we model through Eq.(\ref{total}), the later we refer to it as ``little rip'' phantom matter as in GR it would lead to this kind of event. Then, the modified Friedmann equation (\ref{DGPGB}) can be further expressed as
\begin{align}
E(a)^2=&\Omega_{m0}a^{-3}+\Omega_{d0}\left(\frac{\sqrt{3}}2\frac{\kappa_4A}{H_0\sqrt{\Omega_{d0}}}\ln{a}+1\right)^2\notag\\
       &+2\sqrt{\Omega_{r_c}}\left[1+\Omega_{\alpha}E(a)^2\right]E(a),\label{E(a)}
\end{align}
where $E(a)\equiv H/H_0$, and the dimensionless density parameters are defined as follows:
\begin{align}
&\Omega_{m0}\equiv\frac{\kappa_4^2\rho_{m0}}{3H_0^2}\,\,,&&\Omega_{d0}\equiv\frac{\kappa_4^2\rho_{d0}}{3H_0^2}\,\,,\notag\\
&\Omega_{r_c}\equiv\frac1{4r^2_cH^2_0}\,\,,              &&\Omega_{\alpha}\equiv\frac83\alpha H^2_0\,\,.
\end{align}
It can be easily checked that the modified Friedmann equation (\ref{E(a)}) evaluated at present time $a_0=1$ gives a constraint for these density parameters:
\begin{equation}
\Omega_{m0}+\Omega_{d0}+2\sqrt{\Omega_{r_c}}(1+\Omega_{\alpha})=1.\label{constraint}
\end{equation}
Furthermore, the dimensionless parameters defined in the Eqs.(\ref{barX})-(\ref{rhobar}) can be rewritten in terms of these density parameters (with $\Lambda_5=0$, $\lambda=0$ and $\gamma=1$) as:
\begin{align}
\bar{X}=&\frac{8}3\frac{\alpha}{r_c}H=2\sqrt{\Omega_{r_c}}\Omega_{\alpha}E(a) \label{barX2},\\
b=&\frac{8}3\frac{\alpha}{r_c^2}=4\Omega_{\alpha}\Omega_{r_c}\label{b2},\\
\bar{\rho}=&\frac{32}{27}\frac{\alpha^2\kappa^2_5}{r_c^3}\rho \notag\\
=&4\Omega_{r_c}\Omega_{\alpha}^2\left[\Omega_{m0}a^{-3}+\Omega_{d0}
\left(\frac{3A}{2\sqrt{\rho_{d0}}}\ln{a}+1\right)^2\right]\label{rhobar2}.
\end{align}
%%%%%%%%%%%%%%%%%%%%%%%%%%%%%%%%%%%%%%%%%%%%%%%
\begin{figure*}[!ht]
  \centering
  \subfloat[]{\label{weffrc}\includegraphics[width=0.393\textwidth]{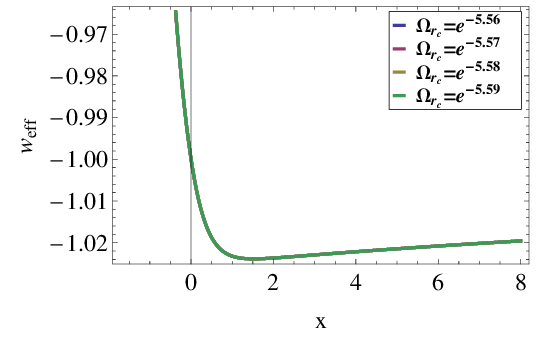}}\qquad~
  \subfloat[]{\label{weffA}\includegraphics[width=0.4\textwidth]{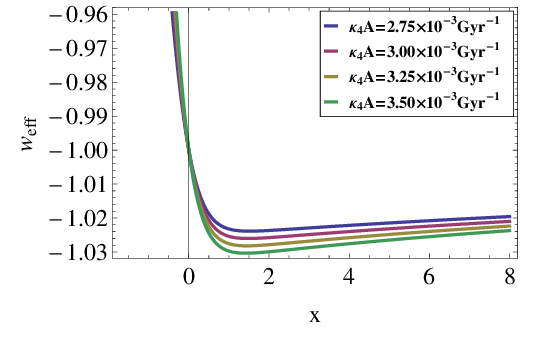}}
  \caption{The effective equation of state (\ref{weffctive}) of the effective dark energy (\ref{DEeff}) against the scalar factor, more precisely $x\equiv\ln{a}$. In figure (a), we have fixed the parameter $\kappa_4 A=2.75\times10^{-3}\textrm{Gyr}^{-1}$, and changed the density parameter $\Omega_{r_c}$ as shown on the plot. In figure (b), we have fixed the density parameter $\Omega_{r_c}=e^{-5.56}$, and changed the parameter $A$ as shown on the plot. We see that the effective dark energy for the model given in Eq.(\ref{E(a)}) will induce a mimicry of a crossing of the phantom divide, i.e., $w$ cross the value $-1$.}
  \label{weff}
\end{figure*}
%%%%%%%%%%%%%%%%%%%%%%%%%%%%%%%%%%%%%%%%%%%%%%%%
%%%%%%%%%%%%%%%%%%%%%%%%%%%%%%%%%%%%%%%%%%%%%%%
\begin{figure*}[!ht]
  \centering
  \subfloat[]{\label{qrc}\includegraphics[width=0.413\textwidth]{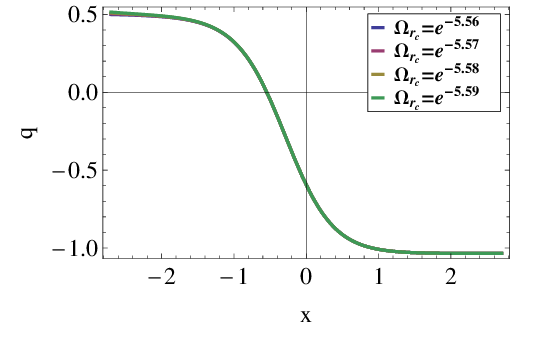}}\qquad~
  \subfloat[]{\label{qA}\includegraphics[width=0.41\textwidth]{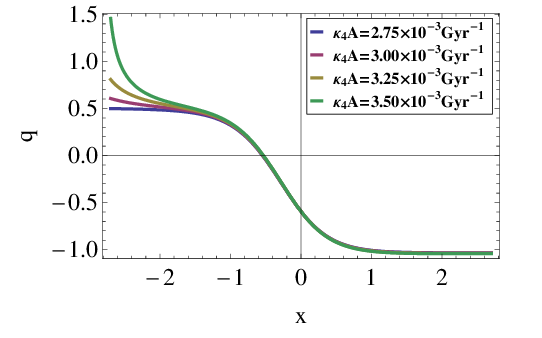}}
  \caption{The deceleration parameter $q$ versus the scale factor, more precisely  $x\equiv\ln{a}$. In figure (a), we have fixed the parameter $\kappa_4 A=2.75\times10^{-3}\textrm{Gyr}^{-1}$, and changed the density parameter $\Omega_{r_c}$ as shown on the plot. As an example, in figure (b), we have fixed the density parameter $\Omega_{r_c}=e^{-5.56}$, and changed the parameter $A$ as shown on the plot.}
  \label{q}
\end{figure*}
%%%%%%%%%%%%%%%%%%%%%%%%%%%%%%%%%%%%%%%%%%%%%%%%

The cosmological evolution of the brane would be equivalent to that of a relativistic model described by GR, i.e.,
\begin{equation}
E(a)^2=\Omega_{m0} a^{-3}+\Omega_{\textrm{eff}},
\end{equation}
with an effective dark energy component $\Omega_{\textrm{eff}}$ defined as follows:
\begin{align}
\Omega_{\textrm{eff}}&\equiv\frac{\kappa_4^2}{3 H_0^2}\rho_{\textrm{eff}}=\Omega_{d0}\left(\frac{\sqrt{3}}2\frac{\kappa_4A}{H_0\sqrt{\Omega_{d0}}}\ln{a}+1\right)^2\notag\\
       &+2\sqrt{\Omega_{r_c}}\left[1+\Omega_{\alpha}E(a)^2\right]E(a).\label{DEeff}
\end{align}
The effective dark energy part consists of the phantom dark energy as well as the higher-dimensional brane-world effect as shown in Eq.(\ref{DEeff}). Furthermore, combining the conservation equation of the effective dark energy:
\begin{equation}
\dot{\rho}_{\textrm{eff}}+3H\rho_{\textrm{eff}}(1+w_{\textrm{eff}})=0,
\end{equation}
with the effective energy density definition (\ref{DEeff}), we obtain the effective equation of state $w_{\textrm{eff}}$:
\begin{equation}
w_{\textrm{eff}}=-1-\frac13\frac{d\ln{\rho_{\textrm{eff}}}}{dx},\label{weffctive}
\end{equation}
where $x\equiv\ln a$. This effective equation of state describes the evolution property of the effective dark energy $\rho_{\textrm{eff}}$, in particular  at late-time.

In addition, the deceleration parameter $q$ is also an important cosmological parameter that could be used to characterize the late-time behavior of the universe and to constrain the model. The deceleration parameter is defined by $q\equiv-\ddot{a}a/\dot{a}^2$, which can be rewritten as
\begin{equation}
q=-1-\frac1{E(a)}\frac{dE(a)}{dx},\label{deceleration}
\end{equation}
where the self-accelerating branch $E(a)$ is given by Eqs.(\ref{barX2}) and (\ref{selfaccelerating}).

Before calculating the effective equation of state $w_{\textrm{eff}}$ and the deceleration parameter $q$ numerically, we highlight that both $w_{\textrm{eff}}$ and $q$ approach negative infinite values when the universe is close to the sudden singularity $\bar \rho\rightarrow \bar \rho_1$, since the cosmic time derivative of the Hubble parameter blows up at the singularity; while the Hubble parameter and the total energy density remain finite.

\subsection{Observational constraints of the model}

%%%%%%%%%%%%%%%%%%%%%%%%%%%%%%%%%%%%%%%%%%%%%%%
\begin{figure*}[!ht]
  \centering
  \subfloat[]{\label{sudsingx}\includegraphics[width=0.41\textwidth]{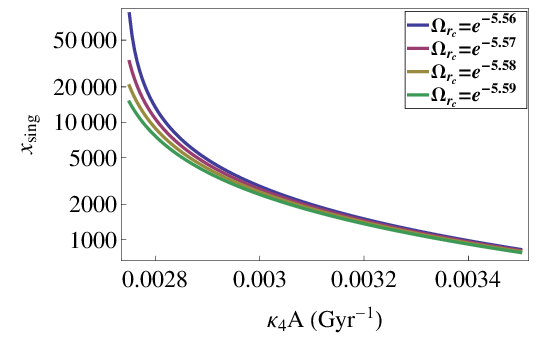}}\qquad~
  \subfloat[]{\label{sudsingt}\includegraphics[width=0.41\textwidth]{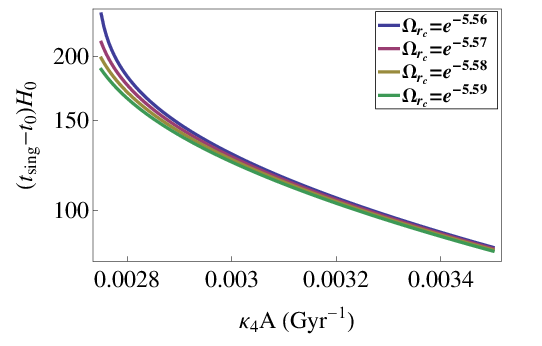}}
  \caption{In figure (a), it shows the value of the scale factor at the sudden singularity $x_{\textrm{sing}}\equiv\ln{a_{\textrm{sing}}}$ versus the parameter $A$ for different values of the density parameter $\Omega_{r_c}$. Here we have fixed the density parameter for matter $\Omega_{m0}=0.27$; while in figure (b), it shows the cosmic time interval before reaching the sudden singularity versus the parameter $A$ for different values of the density parameter $\Omega_{r_c}$.}
  \label{sudsing}
\end{figure*}
%%%%%%%%%%%%%%%%%%%%%%%%%%%%%%%%%%%%%%%%%%%%%%%%
We assume that this self-accelerating DGP-GB model (\ref{E(a)}) behaves much like the standard $\Lambda$CDM model at present and on the near future. Thus we expect the density parameters $\Omega_{r_c}$ and $\Omega_{\alpha}$ due to the brane-world effect to be tiny compared to the other density parameters. Therefore, we could pick out the results derived from the $\Lambda$CDM model to obtain some preliminary constrains of our model; We assume also that the density parameter for matter nowadays $\Omega_{m0}$ to be independent of the cosmological model, i.e., we fix the parameter $\Omega_{m0}=0.27$ for the non-relativistic matter on the brane. Furthermore, we constrain the model by using the deceleration parameter $q_0$ of the $\Lambda$CDM model at the present time: $q_0=-1+3/2 \;\Omega_{m0}=0.595$, which is valid for GR.

We can calculate the deceleration parameter using Eqs.(\ref{deceleration}) and (\ref{E(a)}), which at present reads
\begin{equation}
1+q_0=\frac{3\Omega_{m0}H_0-\kappa_4A\sqrt{3\Omega_{d0}}}{2H_0(1-\sqrt{\Omega_{r_c}}-3\Omega_{\alpha}\sqrt{\Omega_{r_c}})}\,\,.
\end{equation}
The previous relation gives a constraint among the model parameters, which reduces the number of free parameters to two (e.g. $\Omega_{r_c}$ and $A$). In addition, since we expect the model to mimic $\Lambda$CDM cosmology reasonably well at the present time, the best-fit range for the parameter $A$ is considered as obtained in the Ref.~\cite{Frampton:2011sp}: $-2.74\times10^{-3}\textrm{Gyr}^{-1}\leq \kappa_4A\leq 9.67\times10^{-3}\textrm{Gyr}^{-1}$. Despite that negative values of $A$ are allowed observationally, we stick to positive values of $A$, such that the little rip event takes place in GR for a FLRW universe filled with the ``little rip'' phantom matter given in Eq.(\ref{littlerip}).

After imposing the constraints on these parameters, we then evaluate the effective equation of state $w_{\textrm{eff}}$ (\ref{weffctive}) as a function of the scale factor numerically. Notice that the parameter space is picked out such that the density parameter $\Omega_{r_c}$ and $\Omega_{\alpha}$ due to the brane-world effect are very small, and therefore the model behaves almost like the $\Lambda$CDM model at the present time. Fig.~\ref{weff} shows that the behavior of the effective equation of state $w_{\textrm{eff}}$ with respect to the scale factor. We can see that the effects from the DGP-GB brane-world with phantom energy (\ref{littlerip}) will induce a mimicry of a crossing of the phantom divide at the late-time universe (see Fig.~\ref{weff}). Furthermore, it is more sensitive to the parameter $A$ than to the parameter $\Omega_{r_c}$.

In addition, we also calculate and plot the deceleration parameter $q$ [see Eq.(\ref{deceleration})] numerically as shown in Fig.~\ref{q}. This figure shows that our universe has entered in an accelerating expansion phase ($q<0$) recently, roughly since the scale factor reached the value $a\sim0.6$. Besides, the four curves shown in Fig.~\ref{qA}, corresponding to different values of $A$, shows that the deceleration parameter is only sensitive to different values of $A$ in the near past when $a\sim0.14$.

\subsection{Onset of sudden singularity}

As we have seen in the previous section the little rip event in GR can be replaced by a sudden singularity through the self-accelerating branch of the DGP-GB brane-world. We can interpret this result as a smoothing of the little rip in the sense that the Hubble parameter $H$ is finite, even though the sudden singularity happens before the little rip at a finite cosmic time. We further estimate when the sudden singularity will happen in the future within this DGP-GB brane-world, with a cosmological expansion dictated by Eq.~(\ref{E(a)}) and a matter content including a phantom dark energy given in Eq.~(\ref{littlerip}).

The sudden singularity takes place when the dimensionless energy density $\bar\rho\rightarrow\bar\rho_1$ [see Fig.~\ref{branches} and Eq.(\ref{Hdot})]. Therefore, in order to obtain when the sudden singularity is reached, we can directly equate the dimensionless energy parameter $\bar\rho$ (\ref{rhobar2}) to the value of the singularity point $\bar\rho_1$ (\ref{rho1}), and the resulting equation gives the scale factor $x_{\textrm{sing}}\equiv\ln{a_{\textrm{sing}}}$ where the sudden singularity happens. Such a scale factor,  $x_{\textrm{sing}}=\ln{a_{\textrm{sing}}}$, is shown in Fig.~\ref{sudsingx} for various parameter $A$ with different density parameters $\Omega_{r_c}$. We see as well in Fig.~\ref{sudsingx} that the larger the parameter $A$ the sooner the sudden singularity happens. This outcome is reasonable since a larger parameter $A$ generates a smaller effective equation of state $w_{\textrm{eff}}$, i.e., a more negative one, as shown in Fig.~\ref{weffA}, which will cause a sooner doomsday.

Likewise, we can calculate as well the cosmic time remaining from now till the doomsday. This time scale can be obtained by computing the following integration numerically:
\begin{equation}
(t_{\textrm{sing}}-t_0)H_0=\int_{0}^{x_{\textrm{sing}}}\frac{dx}{E(a)}\,\,,
\end{equation}
where $x\equiv\ln{a}$ and the self-accelerating branch $E(a)$ is given by the Eqs.(\ref{barX2}) and (\ref{selfaccelerating}). The numerical results are shown in Fig.~\ref{sudsingt}, where we can compare our results with the age of the universe since the age of our universe at present is roughly of the order of the Hubble time $H_0^{-1}$. Thus as we can see in Fig.~\ref{sudsingt}, the sudden singularity would take place roughly in the far future at a cosmic time of the order of a hundred times the Hubble time, i.e., roughly 1300Gyr.

\section{Conclusion}
The little rip event in a relativistic FLRW universe described by GR can avoid the future space-time singularity at a finite cosmic time; however, it will ultimately lead to the dissolution of all bound structures in the universe. Therefore, it is of interest to investigate if this abrupt event could be smoothed or replaced in some way. Since this kind of doomsday, which is induced by phantom dark energy, as the one given by Eq.~(\ref{littlerip}), will take place at high energy, where UV effect could be required, and in the future, where IR effect could be as well required, we suppose that the combination of UV as well as the IR modifications to GR could give a resolution to this abrupt event. In this respect, we consider here our universe to be described by the self-accelerating branch (\ref{selfaccelerating}) of a generalized brane-world system that is modified by GB modification in the bulk and IG effects on the brane, where the GB/IG effects become significant at the high/late-time regime, respectively. This framework may provide us some insights on the avoidance of abrupt events in the future evolution of the universe.

With the same kind of phantom fluid (\ref{littlerip}) confined on the brane, we discover that a sudden singularity will take place in the future instead of a little rip event. Thus this finding shows that the little rip event is smoothed through a brane-world system with GB and IG effects as we expected, in the sense that the Hubble parameter remains finite during the whole evolution of the brane; nevertheless, the trade-off is that the sudden singularity will happen sooner than the little rip (but still in the far future), where the cosmic time derivative of the Hubble rate will diverge. Based on this analysis, we then study the late-time behavior of the brane and estimate when the sudden singularity will take place in the future within a self-accelerating branch of a DGP-GB model, under the assumption that the self-accelerating DGP-GB brane is filled with dark matter and phantom energy as the one given in Eq.(\ref{littlerip}). We assume as well that the model do not deviate too much from the $\Lambda$CDM model nowadays. Consequently, we conclude that the sudden singularity will occur in a future time roughly of the order of a hundred times the age of the universe, i.e., $\sim$1300Gyr. Moreover, a smaller parameter $A$ in Eq.(\ref{littlerip}) will result in a larger effective equation of state $w_{\textrm{eff}}$ for the effective dark energy $\rho_{\textrm{eff}}$ (\ref{DEeff}), which will push the sudden singularity to a farther future (see Figs.~\ref{weff} and \ref{sudsing}).

Finally, we point out, as demonstrated by our analysis, that in Ref.~\cite{Belkacemi:2011zk} one cannot remove nor appease the occurrence of the little rip singularity within a DGP-GB setup; this is not because of the singularity itself but because of the matter that induces it. In that case, the singularity was caused by a holographic Ricci dark energy component \cite{Belkacemi:2011zk}.

\acknowledgments

M.B.L. is supported by the Basque Foundation for Science IKERBASQUE.
She also wishes to acknowledge the hospitality of LeCosPA Center at the National Taiwan University during the completion of part of this work and the support of the Portuguese Agency ``Funda\c{c}\~{a}o para a Ci\^{e}ncia e Tecnologia" through PTDC/FIS/111032/2009.
P.C. and Y.W.L. are supported by Taiwan National Science Council under Project No. NSC 97-2112-M-002-026-MY3 and by Taiwan’s National Center for Theoretical Sciences (NCTS). P.C. is in addition supported by US Department of Energy under Contract No. DE-AC03-76SF00515.
This work has been supported by a Spanish-Taiwanese Interchange Program with reference 2011TW0010 (Spain) and NSC 101-2923-M-002-006-MY3 (Taiwan)


\begin{thebibliography}{99}
\bibitem{footnote2}
Notice that we have excluded the limiting case where $\mu^2=1/4\alpha$ corresponding to the Chern-Simons gravity, because in that case a homogenous and isotropic brane cannot be embedded in the bulk \cite{Charmousis:2002rc}.
\end{thebibliography}
\end{document}